\documentclass[conference]{IEEEtran}

\usepackage[draft]{hyperref}
\usepackage[dvipsnames]{xcolor}
\usepackage{pgfgantt}
\usepackage{enumitem}
\usepackage{listings}
\usepackage{tikz}
\usepackage{rotating}
\usepackage{glossaries}
\usepackage{graphicx}
\usepackage{makecell}
\usepackage{microtype}
\usepackage{float}
\usepackage{cases}
\usepackage{amsmath}
\usepackage[capitalise,noabbrev]{cleveref}
\usepackage{breakurl}
\usepackage{bookmark}

\def\fnu #1{\footnote{\url{#1}}}
\newcommand{\ready}{Ready}
\newcommand{\unready}{Unready}
\newcommand{\started}{Started}
\newcommand{\healthy}{Healthy}
\newcommand{\unhealthy}{Unhealthy}

\newcommand{\ms}[1]{#1 service} 
\newcommand{\con}[1]{#1 container} 
\newcommand{\ca}[1]{#1 container application} 
\newcommand{\pd}[1]{#1 Pod} 
\newcommand{\prog}[1]{#1 program} 
\newcommand{\ts}[1]{\textbf{\textit{#1}}} 
\newcommand{\cmmnt}[1]{\ignorespaces} 

\usepackage[disable]{todonotes}
\newcommand{\jrcom}[1]{\todo[inline, caption={}, color=blue!40]{JR: #1}}
\newcommand{\ptcom}[1]{\todo[inline, caption={}, color=green!40]{PT: #1}}

\newif\ifhighlightedits
\highlighteditsfalse

\title{Signalling Health for Improved Kubernetes Microservice Availability}

\author{\IEEEauthorblockN{Jacob Roberts, Blair Archibald, Phil Trinder}
\IEEEauthorblockA{\textit{University of Glasgow} \\
Glasgow, Scotland \\
j.roberts.4@research.gla.ac.uk,\{blair.archibald,phil.trinder\}@glasgow.ac.uk}
}

\begin{document}

\maketitle

\begin{abstract}
Microservices are often deployed and managed by a container orchestrator that can detect and fix failures to maintain the service availability critical in many applications.
In Poll-based Container Monitoring (PCM), the orchestrator periodically checks container health.
While a common approach, PCM requires careful tuning, may degrade service availability, and can be slow to detect container health changes.
An alternative is Signal-based Container Monitoring (SCM), where the container signals the orchestrator when its status changes.

We present the design, implementation, and evaluation of an SCM approach for Kubernetes and empirically show that it has benefits over PCM, as predicted by a new mathematical model.
We compare the service availability of SCM and PCM over six experiments using the SockShop benchmark.
SCM does not require that polling intervals are tuned, and yet detects container failure 86\% faster than PCM and container readiness in a comparable time with limited resource overheads.
We find PCM can erroneously detect failures, and this reduces service availability by 4\%.
We propose that orchestrators offer SCM features for faster failure detection than PCM without erroneous detections or careful tuning.
\end{abstract}


\section{Introduction}
\label{sect:intro}

Microservices are a widespread software architecture where an application is factored into small, separate services \cite{microservices}.
It is often critical to maximise service availability; e.g. a retailer will lose sales if a recommender service is unavailable.
Previous work has categorised \cite{silva_microservice_fault_taxonomy,zhou_fault_survey} and investigated predicting microservice failures \cite{zhou_error_prediction}, but there has been little work improving the time to detect failures.

\par

Microservices are often deployed as sets of load-balanced containers, and \textbf{container orchestrators} like \ Kubernetes \cite{kubernetes}, Marathon \cite{marathon}, or Azure Service Fabric \cite{azure_sf} manage the deployment, monitoring, and configuration of container sets~\cite{casalicchio_container_orchestration}.
\cmmnt{Container} Orchestrators monitor container states and take action if the state is undesirable, e.g.\ Kubernetes restarts failed containers\cmmnt{\footnote{Other tools may monitor additional container aspects, e.g. Prometheus \cite{prometheus} can alert on bespoke metrics.}}.
Most \cmmnt{container} orchestrators use \textbf{Poll-based Container Monitoring} (\textbf{PCM}), i.e.\ they periodically interrogate containers to determine their state.
For example, Kubernetes periodically runs liveness probes to determine if a container has failed.
Probes must be carefully tuned to maximise service availability given application characteristics.
Infrequent probes mean a container will have failed for longer before restarting.
In contrast, frequent probes may increase CPU and memory use or incorrectly detect failure due to a temporary increase in response latency.
The need for fast failure detection while avoiding careful tuning makes an alternative \cmmnt{container} monitoring strategy attractive.

\par

In \textbf{Signal-based Container Monitoring} (\textbf{SCM}), a container sends a signal to an orchestrator \cmmnt{or monitoring tool} when it changes state, e.g. when the container is ready to serve requests.
It is reasonable to assume that signal latency is significantly lower than a polling interval, so SCM detects liveness and readiness faster than PCM.
Fast failure detection is crucial for high service availability, e.g.\ in streaming, e-commerce, or financial applications.

\par

This paper describes Kubernetes' PCM and outlines the associated challenges, namely delayed liveness and readiness detection, the requirement for application and hardware-specific tuning of polling frequencies, and the potential for unnecessary container restarts (\cref{sub:kubernetes_ft}).
We then make the following research contributions.

\begin{itemize}[noitemsep]
    \item We describe Signal-based Container Monitoring (SCM) and outline its potential benefits for service availability over PCM, namely faster readiness and liveness detection without careful tuning (\cref{sect:scm}). We develop simple mathematical models for SCM and PCM liveness and readiness detection times and validate them empirically.
    \item We survey the monitoring of six container orchestrators (\cref{sect:other_container_orchestrators}) and find that only one \cmmnt{orchestrator} offers a form of SCM but requires \cmmnt{orchestrator-specific} modified containers. Three orchestrators probe containers as soon as they start, and we give an example where this approach leads to container failure.
    \item We present the design and implementation of a Signal-based Container Monitor for Kubernetes (SCMK), \ts{SKI} (\cref{sect:scmk}).
    \item We present the first ever comparison of poll-based and signal-based monitoring for microservices. Specifically, we compare the service availability of the SockShop benchmark~\cite{sockshop} under Kubernetes with probes and under \ts{SKI} when a container fails. We consider both the default SockShop probes (\ts{DP}) and tuned fast probes (\ts{FP}).
    \begin{itemize}[noitemsep]
        \item Our prototype \ts{SKI} has higher memory and compute overheads than Kubernetes probes. Pod CPU use increases to 0.15 vCPU, 0.04 vCPU (36\%) higher than \ts{FP} and \ts{DP}. The Pod working set increases to 16MiB, 9MiB (129\%) higher than \ts{FP} and \ts{DP} (\cref{sub:overheads}). However, request throughput remains constant under \ts{SKI}, \ts{FP}, and \ts{DP}  (\cref{fig:readiness_throughput,fig:liveness_throughput}).
        \item \ts{SKI} detects container failure in 0.1s, 0.6s (86\%) faster than \ts{FP} and 8.6s (99\%) faster than \ts{DP} (\cref{sub:faster_liveness}).
        \item Both \ts{SKI} and \ts{FP} detect readiness after a failure far faster than \ts{DP} (2.9s and 2.2s compared with $\sim$180s). We propose how the \ts{SKI} implementation could be optimised to reduce the detection time.
        \item FP probes reduce availability by 4\% \cmmnt{to 274 Pods} due to incorrect \cmmnt{container} failure detections (\cref{sub:additional_restarts}), demonstrating the need for careful tuning in PCM.
    \end{itemize}
\end{itemize}

\section{Background}
\label{sect:background}

\subsection{Microservices}

Microservices are popular as they offer advantages over traditional \textbf{monolithic} component-based architectures.
Each service can be horizontally scaled to increase throughput or redundancy, independent of other services.
The services in an application can use the programming language or database technology most appropriate for their function.
Updated software can be released multiple times a day because services can be deployed and updated independently by independent development teams~\cite{monzo_deployment_rate}.

\par

SockShop is a benchmark application that includes RabbitMQ, MongoDB, and MySQL services \cite{sockshop}.
SockShop services are written in Go, NodeJS, Java, and .NET, four of the eight most popular languages used by companies which use containers \cite{datadog_container_report}.
Three \cmmnt{different} services use MongoDB as a database, but these databases are distinct, e.g. the MongoDB service used by the \ms{cart} does not store \ms{user} data.

\par

\begin{figure}[!t]
\centering
\includegraphics[width=\columnwidth]{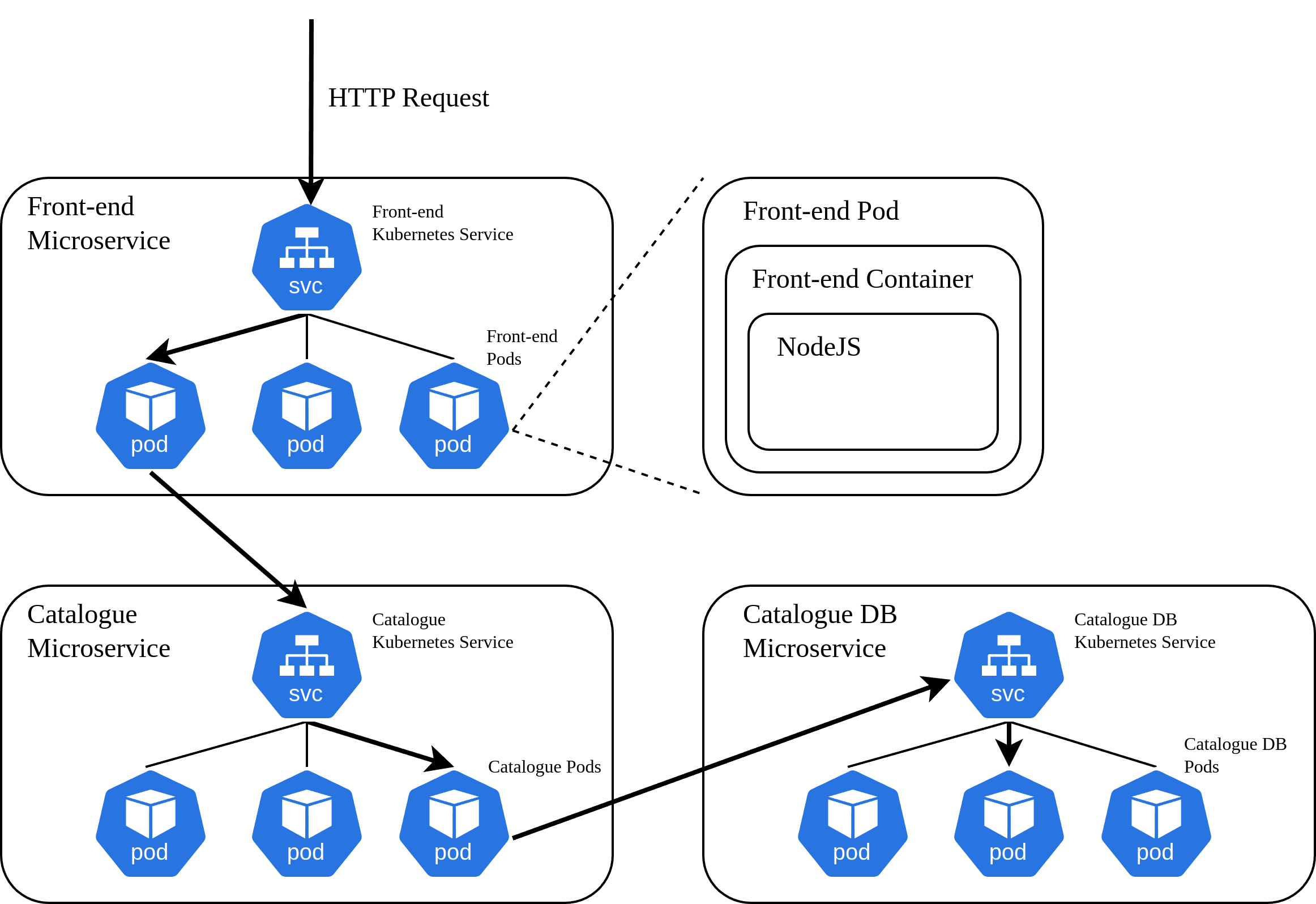}
\caption{Example HTTP request trace to SockShop. Each SockShop service has three service instances (Pods) that may be scheduled on different cluster nodes.}
\label{fig:sockshop_subdiagram}
\end{figure}

SockShop can be deployed as a set of containers using Kubernetes.
For example, a Java \textbf{container application} runs in a \con{cart}, an instance of the \ms{cart}.
A Kubernetes \textbf{Pod} is a group of one or more containers deployed to the same Kubernetes cluster node, e.g. a server or virtual machine.
Containers in a Pod share an IP address, may communicate over a loopback interface, and may share storage volumes.
A \textbf{Kubernetes Service} offers several functions, but here, we view it as a load balancer for a set of Pods.
\cref{fig:sockshop_subdiagram} presents a subset of an example SockShop deployment on Kubernetes and a trace of an HTTP request from a SockShop user.
Each Kubernetes Pod is a service instance.

\par

\par

Some service \cmmnt{faults}failures arise due to asynchronous service interactions \cite{silva_microservice_fault_taxonomy}.
In \textbf{cascading failures}, the failure of one service leads to the failure of another \cmmnt{service} \cite{heorhiadi_gremlin}.
Other \cmmnt{service} failures may be due to logic errors, e.g.~\cite{zhou_fault_survey} gives an example of a database query with an incorrect column name, which may occur in either microservice or monolithic applications.

\par

Although microservice failure rates have been studied, no simple failure distribution models have been found.
Very probably because failures vary greatly with the application and platform.
For example, Sahoo et al. find server failure rate is highly variable over time \cite{sahoo_failure_analysis}.
Garraghan et al. find the mean time between failure of tasks in a production cloud environment is best fit by a Gamma distribution \cite{garraghan_cloud_failure_analysis}.
Over 70\% of surveyed tasks failed within an hour due to user behaviour, skewing the distribution.

\subsection{Kubernetes Fault Tolerance and Challenges}
\label{sub:kubernetes_ft}

\begin{figure}[!t]
\centering
\includegraphics[width=\columnwidth]{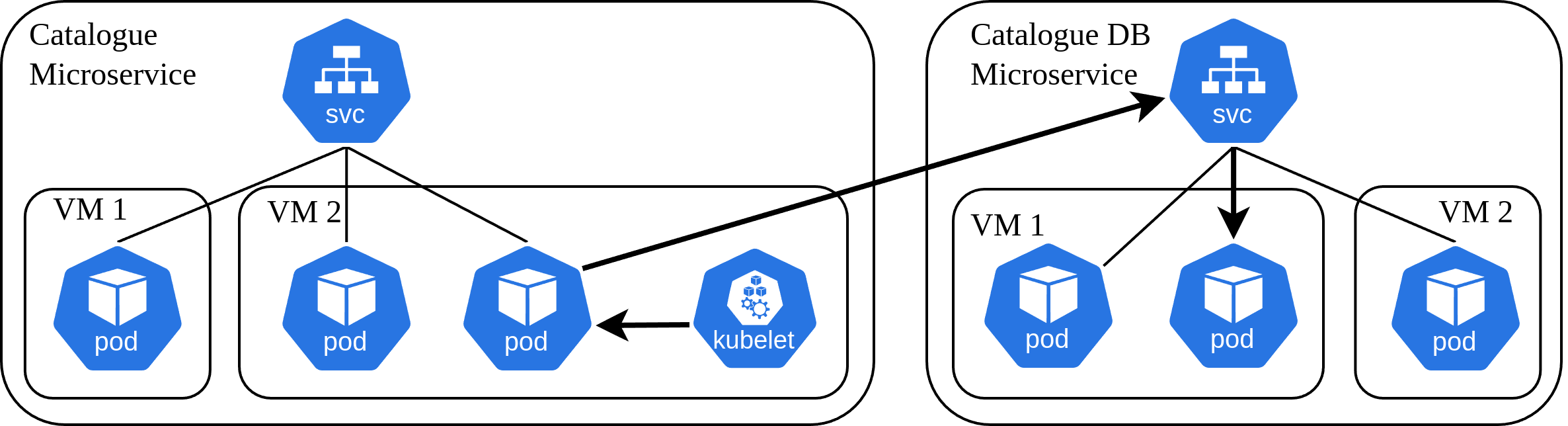}
\caption{Example successful probe communication trace. The kubelet probes a \con{catalogue} container running on the same virtual machine. The container then calls the \ms{catalogue-db}.}
\label{fig:sockshop_probe_subdiagram}
\end{figure}

Kubernetes offers two fault tolerance mechanisms: redundant replica Pods, and restarting "failed" containers.
Each cluster node runs a \textbf{kubelet} component that delegates running containers to a container runtime, e.g.\ CRI-O \cite{crio}, and ensures that Pods scheduled on the node are running.
Kubernetes defines a set of container states.
The kubelet runs \textbf{probes} periodically against the containers on the \cmmnt{cluster} node to determine their state.
Supported probe methods include HTTP requests and gRPC calls.
\cref{fig:sockshop_probe_subdiagram} shows an example trace of a probe\cmmnt{ for the \con{catalogue}}.
A \ca{catalogue} endpoint receives an HTTP request from the local kubelet and pings the \ms{catalogue-db}.
A \con{catalogue-db} running on another Kubernetes \cmmnt{cluster} node may process this ping.
There are three types of probes - liveness, readiness, and startup.
Users may define up to one probe of each type for a container.
If a probe type is not configured for a container, Kubernetes assumes it will succeed.

\par

\par

When a container passes a \textbf{readiness} probe, it is \textbf{\ready{}} to process requests.
A container is \textbf{\unready{}} until it passes a readiness probe.
Once all containers in a Pod are \ready{}, so is the Pod, and a Kubernetes Service will direct requests to it.
A service is unavailable if all associated Pods are \unready{} or \ready{} Pods cannot achieve sufficient throughput.

\par

\textbf{Liveness probes} determine if a container is \textbf{\healthy{}}.
Containers that fail liveness probes or are reported exited by the container runtime are marked \textbf{\unhealthy{}}.
Kubernetes requests the container runtime to restart \unhealthy{} containers.
If a container is restarted too frequently, a \textbf{CrashLoopBackOff} mechanism delays subsequent restarts to mitigate the impact on other containers.

\par

\begin{figure*}[!t]
\centering
\includegraphics[width=\textwidth]{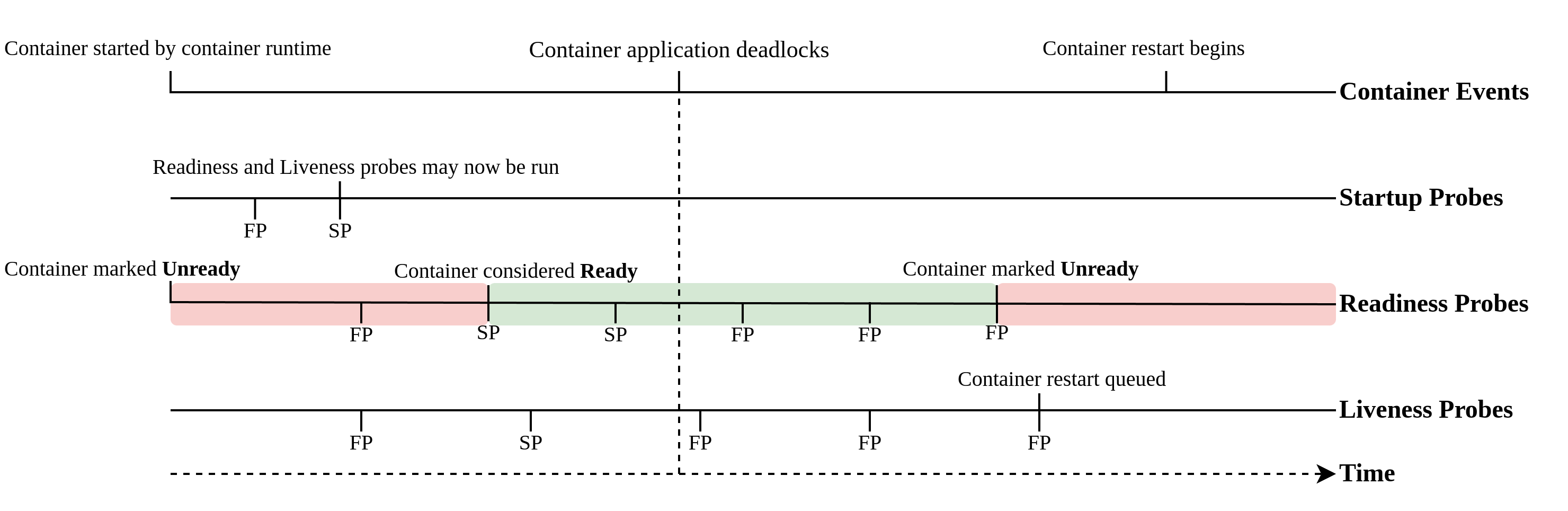}
\caption{An example timeline of failing (FP) and succeeding (SP) startup, readiness, and liveness probes with container failure. Periods where the container \ready{} and \unready{} are shown in green and red.}
\label{fig:probe_timeline}
\end{figure*}

By default, a container is marked \unready{}/\unhealthy{} after three consecutive failed readiness/liveness probes.
An \unready{} container may become \ready{} at a later time, whereas Kubernetes will queue an \unhealthy{} container to be restarted.
Kubernetes treats a container as \unready{} while it is restarting.
\cref{fig:probe_timeline} shows an example timeline where a container application experiences a deadlock and stops responding to probes.
The figure includes \textbf{startup} probes, which are explained later in this section.
In \cref{fig:probe_timeline}, the readiness and liveness probes have different periods and detect the container is \unready{} or \unhealthy{} at different times.

\par

\par

By default, Kubernetes sends a container probes as soon as the container has started, but users can configure probes to begin after an initial delay.
This delay is beneficial when container applications must initialise before serving requests but must be carefully tuned because \emph{container start times depend on the container application and the underlying hardware \cite{straesser_container_start_times}}.
Delaying container failure or readiness detection decreases service availability.
Kubernetes provides \textbf{startup} probes to account for varying \cmmnt{container} initialisation times.
Once a startup probe passes, Kubernetes marks the container \cmmnt{application} as \textbf{\started{}} and stops sending startup probes to the container.
Kubernetes will not execute readiness or liveness probes until a startup probe succeeds, as shown in \cref{fig:probe_timeline}.

\par

The probing interval must also be tuned as probes will, on average, detect failure or readiness in half of the probing interval.
A longer interval leads to slower detections, but \emph{developers may consider whether frequent probes increase the resource use of containers}, e.g. memory \cite{gitlab_kubernetes_fault}.

\par

\emph{Probes may degrade service availability.}
For example, DoorDash reported an incident where readiness probes failed due to increased latency to a service the probed container called \cite{doordash_kubernetes_fault}.
\unready{} containers led to increased requests to the remaining \ready{} containers, high CPU utilisation, and increased response latency.

\par

\paragraph*{Summary of Probe Problems}
Probes require hardware and application-specific tuning and may increase resource usage or degrade service availability.
Because readiness or failure is detected when a probe runs, it may be slow to detect state changes.
These challenges motivate research into alternative monitoring methods.

\section{Container Monitoring Survey}
\label{sect:other_container_orchestrators}

\cref{tab:detection_method_comparison} compares the monitoring methods used in six popular container orchestrators and the SCMK implementation presented in \cref{sect:scmk}, \ts{SKI}.
Whether liveness and readiness are distinguished and the approach to allow containers to initialise are compared.
Orchestrators based on Kubernetes, such as \cite{aws_eks}, \cite{azure_aks}, and \cite{gcp_gke} are not considered.
Some surveyed orchestrators can manage entities other than containers, such as virtual machines.
We only consider orchestrator features pertaining to container monitoring.

\par

We consider the three Azure Service Fabric container monitoring mechanisms: probes, watchdogs, and health reports.
The first two are polling mechanisms (PCM), while the latter is the only signal-based monitoring (SCM) offered by the orchestrators surveyed.
Azure Service Fabric provides liveness and readiness probes similar to Kubernetes but marks newly started containers \ready{} until readiness probes fail.
Microsoft suggests monitoring using a \textbf{watchdog}.
A watchdog monitors other services' health, generates health reports, and may take remedial actions, e.g. periodically deleting log files.
The FabricObserver \cite{fabricobserver} watchdog monitors services through \textbf{Observers}, e.g.\ default Observers poll CPU and RAM use and periodically test if an HTTP endpoint is available.

\par

Containers managed by Azure Service Fabric may also report their state through a \textbf{health report} API that is more granular than probes—for example, reporting disk, RAM, or CPU use in separate health reports, where the state may be 'OK', 'Warning', or 'Error'.
Health reports are stored within Azure Service Fabric, and tools may query them to determine if remedial action is required.

\par

\begin{table*}[!t]
\centering
    \caption{Comparing the monitoring methods in popular orchestrators. We consider the default FabricObserver Observers.}
    \label{tab:detection_method_comparison}
    \begin{tabular}{@{}llll@{}} 
        Orchestrator & \makecell[l]{Monitoring\\Method} & \makecell[l]{Distinguishes Readiness\\From Liveness}& \makecell[l]{Allowance For Container Initialisation} \\
        \hline
        Kubernetes & Probes & Yes & Can delay start of probes \\
        Nomad & Probes & \makecell[l]{At deployment time} & \makecell[l]{Failing readiness checks are tolerated} \\
        Docker Swarm & Probes & No & \makecell[l]{Ignores failing probes for a period} \\
        Marathon & Probes & \makecell[l]{At deployment time} & \makecell[l]{Ignores failing probes for a period} \\
        AWS ECS & Probes & No & \makecell[l]{Ignores failing probes for a period} \\
        Azure Service Fabric (Probes) & Probes & Yes & Can delay start of probes \\
        Azure Service Fabric (FabricObserver) & Probes & N/A & N/A \\
        Azure Service Fabric (Health Reports) & Signal & N/A & N/A \\
        \textbf{SKI} & \textbf{Signal} & \textbf{Yes} & \textbf{N/A} \\
    \end{tabular}
\end{table*}

\par

Kubernetes, Nomad~\cite{nomad}, Marathon~\cite{marathon}, and Azure Service Fabric~\cite{azure_sf} distinguish between readiness and liveness probes.
Conflating \cmmnt{container} readiness and liveness may be detrimental to service availability.
For example, if readiness probes fail while a container has high resource use, it is not sent requests, and resource use can decrease without a container restart \cite{breck_probes_presentation}.
This strategy relies on the distinction between readiness and liveness to avoid restarts.

\par

Docker Swarm~\cite{docker}, Marathon~\cite{marathon}, and AWS ECS~\cite{aws_ecs} use a probing strategy that may decrease service availability.
These orchestrators do not delay the start of probing when a container starts but do not restart a container if probes fail for a period after the container starts.
For example, a user may want to probe a container every 10s after it has taken 60s to initialise.
These orchestrators would probe the container 10s after it starts but not mark it as failed for 60s.
SockShop contains an example where this monitoring behaviour leads to unnecessary restarts.
When deploying SockShop, the \ms{catalogue-db} is initially unavailable.
Probes for \con{catalogue}s fail and prevent subsequent probes from succeeding, causing the orchestrator to restart the \con{catalogue}s.

\section{Signal-based Container Monitoring}
\label{sect:scm}

Under \textbf{Poll-based Container Monitoring} (\textbf{PCM}), the container orchestrator determines a container's state periodically.
Under \textbf{Signal-based Container Monitoring} (\textbf{SCM}), the orchestrator or a monitoring tool receives a signal emitted by a container when its state changes.
For example, a container application might send an HTTP PUT request to an orchestrator when it can serve HTTP requests.
Kubernetes' detecting failure via a container runtime is not SCM as this is not configurable and restricted to only detecting failure.
Unlike PCM, SCM does not require careful tuning, as no polling intervals are configured.
Instead, practitioners identify what signals to use.
Under PCM, the orchestrator detects the container state the next time it polls the container state and waits, on average, half of the polling interval before detecting the container state.
Under SCM, the orchestrator detects the container state when it receives the appropriate signal.
Detecting either container failure or readiness faster increases service availability.

\subsection{Modelling Liveness/Failure Detection}
\label{sub:liveness_model}

A simple model of liveness detection time assumes a signal can be sent upon failure, e.g.\ the container is not deadlocked, and that the failure is detectable by both signals and probes.
For SCM, the mean time to detect failure, or the lack of liveness, $T_{l,s}$ depends only on the signal latency, $L_s$.
\begin{equation}
    T_{l,s} \approx L_s
    \label{eqn:scm_liveness}
\end{equation}

For PCM, the mean time to detect failure, $T_{l,p}$, depends on the liveness probe latency $L_l$, the probe interval $I_l$, and the number of consecutive successful/failed probes required to establish liveness/failure $N_l$.
\begin{equation}
    T_{l,p} \approx (N_l-1/2)I_l + L_l
    \label{eqn:pcm_liveness}
\end{equation}
On average, failure will occur in the middle of the probe interval, and the subsequent liveness probe will fail, hence $N_l-1/2$.
For example, by default, Kubernetes detects failure after 3 consecutive liveness probes fail, so $T_{l,p} \approx 2.5I_l + L_l$.

\par

It is reasonable to assume that $L_s \ll I_l$, e.g. the Kubernetes minimum $I_l$ is 1s and the Sockshop default is 3s.
It is similarly reasonable to assume that $L_s \leq L_l$ because the probe latency includes the time for the container to process a probe. This accords with our measurements of $L_s \approx 0.1s$ for \ts{SKI} and $L_l \approx 0.2s$ for \ts{FP} (\cref{sub:scm_validation}).
Under these assumptions, \cref{eqn:scm_liveness,eqn:pcm_liveness} predict that \emph{SCM detects failure/liveness far faster than PCM}.
The key intuition is that a signal is sent immediately upon failure under SCM while PCM must wait for the relatively long probe interval to elapse. This prediction is borne out by our evaluation in \cref{sub:faster_liveness}.

\subsection{Modelling Readiness Detection}
\label{sub:readiness_model}

The time to detect readiness under SCM or PCM depends on the time for a container to become \ready{} and for monitoring to begin.
This contrasts with liveness detection, which occurs after a container has restarted and monitoring begins.

\par

For SCM, the mean time to detect readiness, $T_{r,s}$, is determined by either the time for the container to become \ready{}, $T_c$, or the time to begin monitoring the container, $T_s$.
Once the longest of these has elapsed, readiness is detected after a signal latency, $L_s$.
\begin{equation}
    T_{r,s} \approx max(T_c,T_s) + L_s
    \label{eqn:scm_start}
\end{equation}
For PCM, the mean time to detect readiness, $T_{r,p}$, depends on \cmmnt{the time for the container to become \ready{},} $T_c$, the time until the first readiness probe, $T_r$, the number of readiness probes required, $N_r$, and readiness probe latency $L_r$.
If the container is \ready{} before the first readiness probe, i.e. $T_c<T_r$, then readiness is detected after a further $N_r-1$ probes succeed \cref{eqn:pcm_start_a}.
If the container becomes \ready{} after the first readiness probe, then the same logic as \cref{eqn:pcm_liveness} applies in \cref{eqn:pcm_start_b}.
If $L_s \ll I_r$ and $L_s \leq L_r$, it is expected that $L_s \leq (N_r-1)I_r + L_r$, but SCM may detect readiness slower than PCM if $T_s > T_r$.
\begin{subnumcases}{T_{r,p} \approx\label{eqn:pcm_start}}
    T_r + (N_r-1)I_r + L_r & \text{if $T_c<T_r$,}\label{eqn:pcm_start_a} \\
    T_c + (N_r-1/2)I_r + L_r & \text{otherwise.}\label{eqn:pcm_start_b}
\end{subnumcases}

Orchestrators that do not distinguish between liveness and readiness (see \cref{sect:other_container_orchestrators}) also follow \cref{eqn:scm_start,eqn:pcm_start}, but failed probes lead to container restarts.

\subsection{Model Validation}
\label{sub:scm_validation}

\cref{tab:probe_settings} presents the probe parameters for the Kubernetes probe configurations used in the readiness and liveness evaluations in \cref{sect:results}.
The configurations are SockShop Default Probes (\ts{DP}) and Fast Probes (\ts{FP}) that are tuned to reduce the time to detect failures and readiness.
We use these values to confirm the validity of the SCM and PCM models.

\begin{table*}[!t]
\centering
    \caption{The Kubernetes probe parameters used in the experiments in \cref{sect:results}, and to validate the SCM and PCM Models in \cref{sect:scm}}
    \label{tab:probe_settings}
    \begin{tabular}{@{}lllll@{}} 
        \makecell[l]{Monitoring Configuration (Probe Type)} & Experiments & \makecell[l]{Number of Probes Required} & \makecell[l]{Probe Interval (s)} & \makecell[l]{Initial Probe Delay (s)}\\
        \hline
        Default Probes (DP) & & & &\\
        Liveness Probes  & All & 3 & 3 & 300\\
        Readiness Probes  & All & 1 & 3 & 180\\
        \hline 
        Fast Probes (FP) & & & &\\
        Liveness Probes  & Liveness & 1 & 1 & 1\\
        Readiness Probes & Readiness & 1 & 1 & 1\\
    \end{tabular}
\end{table*}

\paragraph*{Liveness/Failure Detection}
The second column of \cref{tab:failure_detection_times} in \cref{sub:faster_liveness} reports the time to queue a container restart after failure, taken to be the times to detect failure, $T_{l,s}$ and $T_{l,p}$.
The measured time for \ts{SKI} is $T_{l,s} \approx L_s \approx 0.1s$.
The same column shows $T_{l,p}$ for \ts{FP} is 0.7s, which can be used to compute $L_l$.
Substituting the parameters from \cref{tab:probe_settings} into \cref{eqn:pcm_liveness} gives $T_{l,p} \approx (1-1/2)1s + L_l \approx 0.5s + L_l \approx 0.7s$, so $L_l \approx 0.2s$.
Our model then predicts that the mean time to detect liveness for \ts{DP} is $T_{l,p} \approx (3-1/2)3s + 0.2s \approx 7.7s$.
This is 1s lower than the 8.7s measured in~\cref{tab:failure_detection_times}.
We hypothesise that the difference is because \cref{eqn:pcm_liveness} estimates mean detection time assuming a uniform probability of failure, but the failure in the tests occurs early in a probe interval.

\paragraph*{Readiness Detection}
In \cref{sub:slower_readiness} the container becomes \ready{} before monitoring begins, so $T_c<T_r$ and $T_c<T_s$.
The rightmost column of \cref{tab:readiness_requests} reports the number of timed-out HTTP requests after container failure.
This count multiplied by the 0.5s timeout duration approximates the readiness detection times, $T_{r,s}$ and $T_{r,p}$.
So for \ts{SKI} $T_{r,s} \approx 5.7 \times 0.5 \approx 2.9s$.
From $T_{r,s}$, we can use \cref{eqn:scm_start} to compute the time for \ts{SKI} to begin monitoring the container, $T_s$, as $T_s \approx T_{r,s} - L_s \approx 2.9s - 0.1s \approx 2.8s$.

We follow the same reasoning to compute readiness times for PCM from the failed requests in \cref{tab:readiness_requests}.
For \ts{FP} $T_{r,p} \approx 4.4 \times 0.5 \approx 2.2s$, and for \ts{DP} $T_{r,p} \approx 362 \times 0.5 \approx 181s$.
For \ts{FP} substituting the parameters from \cref{tab:probe_settings} into \cref{eqn:pcm_start_a} gives $T_{r,p} \approx T_r + (1-1)1 + L_r$.
As the readiness and liveness probes call the same endpoint, it is reasonable to assume $L_r \approx L_l \approx 0.2s$, giving $T_{r,p} \approx T_r + 0.2s \approx 2.2s$, and so the time until the first readiness probe, $T_r$, is 2s for \ts{FP}.

For \ts{DP} the measured readiness time $T_{r,p}$ of 181s is consistent with \cref{eqn:pcm_start_a} as the substantial time until the first readiness probe (180s) dominates. That is, substituting the parameters from \cref{tab:probe_settings} into  \cref{eqn:pcm_start_a} gives $T_{r,p} \approx 180 + (1-1)3 + 0.2 \approx 180s$.

\subsection{Discussion}

While simple, these models are consistent with the implementations and show that, under reasonable assumptions, SCM always detects liveness faster than PCM as it does not need to wait for some, or all, of a probe interval to elapse.
More elaborate models could include distributions for latency, container start times, monitoring start times, and other fault distributions/models.

\section{Signal-based Container Monitoring for Kubernetes}
\label{sect:scmk}

\subsection{SKI Architecture}
\label{sub:scmk_prototype}

We present a prototype implementation of Signal-based Container Monitoring for Kubernetes (\textbf{SCMK}), \ts{SKI}.
While \ts{SKI} is simpler to configure than probes as there are no intervals to tune, suitable logs must be identified to provide failure signals.

\par

\ts{SKI} is deployed as a \textbf{sidecar container} in the same Pod as the monitored 'main' container~\cite{kubernetes_toolkit_patterns}.
Sidecar containers typically provide additional functionality, e.g.\ to copy container logs from a shared filesystem to a centralised log analyser.
We extend the kubelet to open a local UNIX socket and listen for messages from \ts{SKI}.

\par

\begin{figure}[!t]
\centering
\includegraphics[width=\columnwidth]{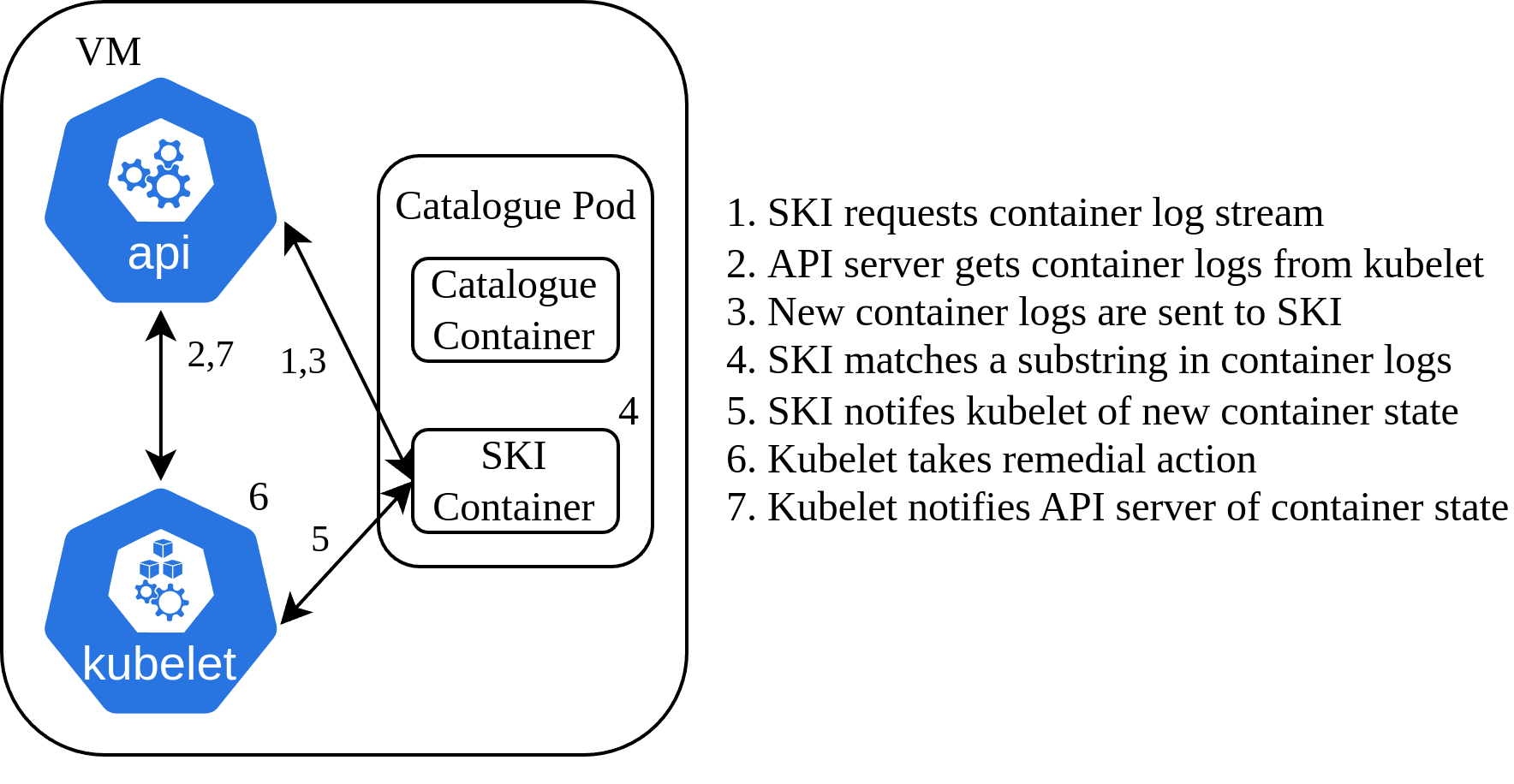}
\caption{Prototype SKI software architecture showing the monitored container and sequence of communications.}
\label{fig:pod_with_supervisor}
\end{figure}

\par

\ts{SKI} fetches container application logs via a Kubernetes API and interprets those that match a given string as signals.
\cref{fig:pod_with_supervisor} shows the software architecture and communications of \ts{SKI}.
When \ts{SKI} receives a matching log, indicating the monitored container is \ready{} or \unhealthy{}, it sends a message to the extended kubelet.
The kubelet then marks the container as \ready{} or queues a \cmmnt{container} restart as appropriate.

\subsection{Alternative Architectures}
\label{sub:scmk_alternatives}

\ts{SKI} is simple to implement and evaluate (\cref{sect:results}), but fewer communications may reduce \cmmnt{signal latency, $L_s$ and} the time to begin monitoring, $T_{r,s}$ in \cref{eqn:scm_start}, e.g. if the monitored container signalled \ts{SKI} via an HTTP PUT request.
This approach requires changes to the container, but this limitation does not necessarily prevent adoption.
For example, the popular poll-based metrics monitor Prometheus~\cite{prometheus} requires applications to use a client library.
A less restrictive approach is to write logs to files on a storage volume within the Pod.
If the volume is shared with a sidecar container \cmmnt{, a common pattern~\cite{kubernetes_toolkit_patterns},} logs can be read without the Kubernetes API, and the container remains orchestrator-agnostic.

\par

Kubernetes and \ts{SKI} do not aggregate the status of multiple Pods, e.g. the total number of \ready{} Pods for a service, before restarting a container.
Other SCMK implementations may improve service availability by performing this aggregation, e.g. marking containers with high CPU usage \unhealthy{} unless half of the containers associated with the service are \unready{}.

\subsection{Comparison to Azure Service Fabric}
\label{sub:scmk_vs_asf}

\cref{sect:other_container_orchestrators} found Azure Service Fabric supported SCM by allowing adapted containers to post health reports.
\ts{SKI} and one of the alternative architectures in \cref{sub:scmk_alternatives} allows monitored containers to remain orchestrator-agnostic.
In another alternative architecture, the container emits an HTTP request and depends on a container-facing API, not Kubernetes.
For example, the container might send the same request to a Signal-based Container Monitoring for Nomad tool.

\section{Comparing Signal and Poll-based Container Monitoring}
\label{sect:results}

\subsection{Evaluation Framework}
The evaluation uses the popular SockShop benchmark application~\cite{sockshop}, and specifically the \ms{catalogue} because it becomes \ready{} quickly ($\sim$2s) after restarting, and  does not require user authentication, simplifying testing.
\cref{fig:sockshop_probe_subdiagram} shows that probes to \con{a catalogue} ping \ms{the catalogue-db}.
\ts{SKI} matches logs that do not require this connectivity.
We believe this does not invalidate results as \ms{the catalogue-db} is available during tests.

\par

To determine when the \ms{catalogue} is available, HTTP requests are generated, throughput measured, and failed requests logged using the popular Locust \cite{locust} load testing framework.
Because Locust is used to determine service availability, not the impact of container monitoring on throughput, we limit throughput to 100 requests/s.

\par

Experiments are executed on a 1-node Kubernetes v1.27.8 cluster using CRI-O v1.27.1 hosted on an Ubuntu 20.04 virtual machine with 32GB RAM and 8 vCPUs\footnote{The \cmmnt{virtual machine} RAM and vCPU AWS recommend for microservices \cite{aws_ms_ec2_recommendation}.}.
\cmmnt{During each test, }SockShop is deployed with 1 Pod per service and allowed 6 minutes for all containers to be \ready{} before Locust, running on the same host, begins generating requests.
Each test configuration is run thirty times.
Times are measured to the nearest second because the shortest permitted probe interval is 1s and containers may take several minutes before they are \ready{}.

\par

Container failure is induced by terminating the HTTP request-handling process, the \prog{catalogue}, 60s after Locust starts.
During readiness detection evaluations, catalogue runs as PID 1 inside the container.
When PID 1 terminates, the container is exited in CRI-O, and Kubernetes marks the container \unhealthy{}.
Probes and \ts{SKI} detect the restarted container becoming \ready{} but not the failure.
During liveness detection evaluations, catalogue executes under a shell running as PID 1.
After catalogue terminates, the shell writes a log and hangs.
Because PID 1 is running but the container cannot serve probes, Kubernetes detects this failure through liveness probes or \ts{SKI} and sends a \texttt{SIGTERM} OS signal to PID 1.
After a termination grace period, 30s by default, a \texttt{SIGKILL} OS signal is sent if PID 1 is still running.

\par

Four monitoring configurations are used to evaluate container readiness and liveness detection.
Readiness (\cref{sub:slower_readiness}) and liveness (\cref{sub:faster_liveness,sub:additional_restarts}) detection are each evaluated using three monitoring configurations. 
\ts{Default Probes} (\ts{DP}), the probes provided with SockShop and \ts{SKI}, where \ts{SKI} is deployed per \cref{fig:pod_with_supervisor}, are used in both readiness and liveness tests.
\ts{SKI} uses a termination grace period of 2s, the minimum hardcoded in the kubelet, and still uses \ts{DP} probes so the \con{catalogue} is not marked \ready{} immediately.
The kubelet logs show that in \ts{SKI} tests \ts{SKI}, not probes, detects readiness or failure.
Different variants of \ts{Fast Probes} (\ts{FP}) are used when evaluating readiness \cmmnt{(\cref{sub:slower_readiness})} and liveness \cmmnt{(\cref{sub:faster_liveness,sub:additional_restarts})} detection.
Both variants reduce the probing interval of \ts{DP} to 1s and use startup probes to allow the container \cmmnt{application} to initialise.
1s is the minimum probing interval Kubernetes allows, \emph{so \ts{FP} should detect readiness as fast as probes allow.}
Under \ts{FP}, the initial probe delay is 1s, though the default is 0s.
If the delay is 0s, we expect the first probe to fail and the results to be the same.
Developers may decide that a probe interval of 1s is too frequent, e.g. developers may consider the impact of readiness probes on memory usage \cite{gitlab_kubernetes_fault} or probes may take longer than 1s to run \cite{gitlab_kubernetes_merge}.
\emph{During liveness evaluations \ts{FP} detects failures and begin restarts as fast as allowed} by also using a 2s grace period and marking \con{the catalogue} \unhealthy{} after one failed liveness probe instead of the default three.
\emph{The liveness \ts{FP} configuration is not suitable for readiness evaluations due to additional restarts (\cref{sub:additional_restarts}), demonstrating the need to tune PCM carefully.}

\subsection{Overheads}
\label{sub:overheads}

When the monitored container failed\cmmnt{ during evaluation}, the working set of the Pod was 16 MiB under \ts{SKI}, 9 MiB (129\%) higher than under \ts{FP} and \ts{DP} probes.
At the same time, Pod CPU use was 0.15 vCPU under \ts{SKI}, 0.04 vCPU (36\%) higher than under \ts{FP} and \ts{DP}.
Alternative SCMK architectures may reduce the monitoring overhead, e.g., where one \ts{SKI} container monitors \cmmnt{containers in} multiple Pods.
Throughput of the monitored service was constant across \ts{SKI}, \ts{DP}, and \ts{FP} (\cref{fig:readiness_throughput,fig:liveness_throughput}).

\subsection{Detecting Readiness}
\label{sub:slower_readiness}

\begin{figure*}[!t]
\centering
\includegraphics[width=\textwidth]{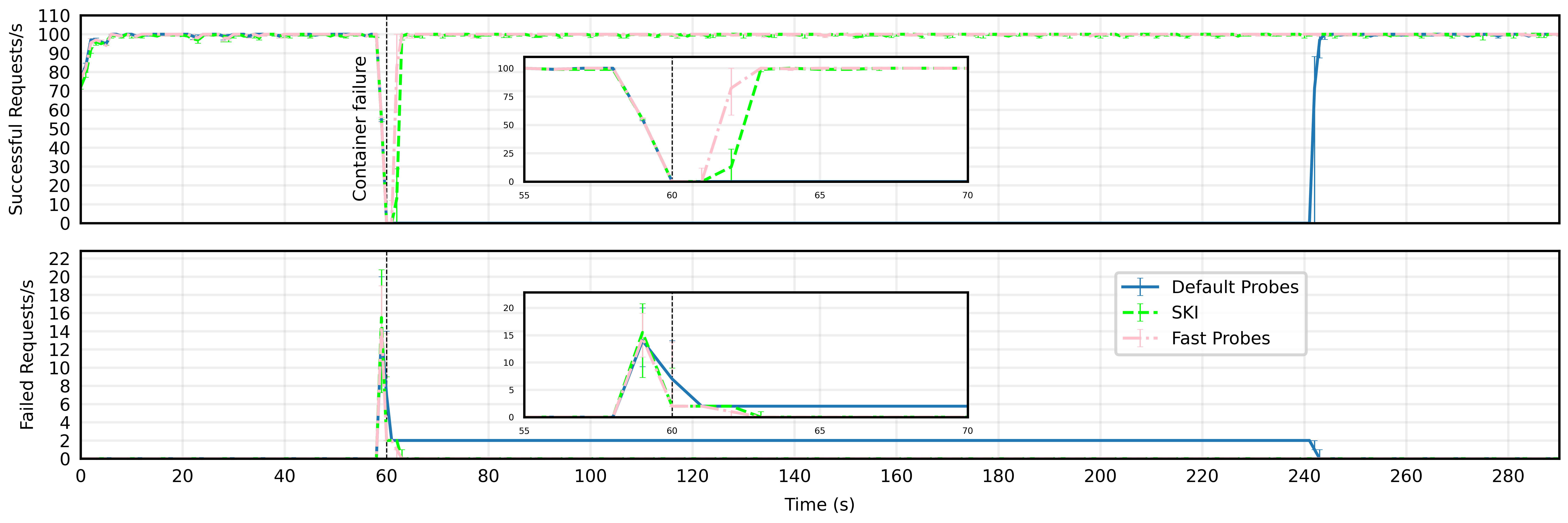}
\caption{The median successful and failed requests, with upper and lower quartiles, of the \ms{catalogue} under default probes (blue), SKI (green), and fast probes (pink) when evaluating readiness. The container recovers in 5s under SKI and fast probes.}
\label{fig:readiness_throughput}
\end{figure*}

\cref{fig:readiness_throughput} plots the throughput and failed requests in requests/s of \ms{the catalogue} over 290s with a container failure at 60s.
\emph{\ts{DP} detects readiness after $\sim$180s, the initial probe delay in \cref{tab:probe_settings}, and throughput returns to 100 requests/s at 240s, whereas throughput recovers after 5s under \ts{FP} and \ts{SKI}.}

\par

When the container fails, the median rate of failed requests peaks and the interquartile range of all three configurations contains the three medians, indicating that \emph{the initial rate of failed requests is independent of monitoring configuration when CRI-O detects failure}.
The plateau of 2 requests/s (2\%) is because the HTTP requests have a 0.5s timeout.

\par

\par

To quantify the time to detect readiness, described by \cref{eqn:scm_start,eqn:pcm_start}, \cref{tab:readiness_requests} presents the total requests sent, the total failed, and those that failed due to timeout.
As the request timeout is 0.5s, \cmmnt{we estimate that} \emph{\ts{SKI} detects readiness 2.9s after container failure, 0.7s slower than \ts{FP} which we attribute to the \ts{SKI} architecture.}
When a monitored container fails \ts{SKI} waits for 1s, polls Kubernetes until it reports that the container has restarted, waits for 1s, and then requests the container's logs.
Other SCMK implementations may begin container monitoring and detect readiness faster if they do not use Kubernetes to fetch logs or detect the container restart.

\par

\begin{table}[!t]
\centering
    \caption{Measuring time to detect \con{catalogue} readiness after container failure via timed-out HTTP requests. Mean and standard deviation of the total and failed requests to \ms{catalogue}, and those that failed due to timeout.}
    \label{tab:readiness_requests}
    \begin{tabular}{@{}llll@{}} 
        \makecell[l]{Monitoring\\Configuration} & \makecell[l]{Total \\Requests} & \makecell[l]{Total Failed\\Requests} & \makecell[l]{Timed Out\\Requests} \\
        \hline
        Default Probes & $11100\pm100$ & $380\pm10$ & $362\hphantom{.0}\pm2$ \\
        Fast Probes & $28600\pm30$ & $\hphantom{3}20\pm10$ & $\hphantom{36}4.4\pm0.5$ \\
        \makecell[l]{SKI} & $28450\pm20$ & $\hphantom{3}30\pm10$ & $\hphantom{36}5.7\pm0.5$ \\
    \end{tabular}
\end{table}

\par

\paragraph*{Summary}
Throughput recovers to $\sim$100 requests/s in less than 5s under \ts{FP} and \ts{SKI}, significantly faster than the $\sim$180s under \ts{DP}.
The massive reduction in failure recovery time for \ts{FP} compared with \ts{DP} demonstrates why tuning is highly desirable under PCM.
\ts{SKI} detects readiness in 2.9s, 0.7s slower than \ts{FP}, although this could be improved by optimising the \ts{SKI} architecture.

\subsection{Detecting Failure}
\label{sub:faster_liveness}

\cref{tab:failure_detection_times} measures time to detect container failure by reporting the mean time for Kubernetes to queue and begin a container restart, both logged by the kubelet, after the container fails.
The time to queue the restart is the time to detect failure described by \cref{eqn:scm_liveness,eqn:pcm_liveness}.
We omit three \ts{FP} outliers as CrashLoopBackOff delayed the restarts.
\emph{\ts{SKI} detects failure in 0.1s, 0.6s (86\%) faster than \ts{FP} and 8.6s (99\%) faster than \ts{DP}}.

\par

\begin{table}[!t]
\centering
    \caption{Measuring time to detect \con{catalogue} failure. Mean and standard deviation of the time to queue and restart \con{the catalogue} after termination.}
    \label{tab:failure_detection_times}
    \begin{tabular}{@{}llll@{}}
        \makecell[l]{Monitoring\\Configuration} & \makecell[l]{Time to queue\\container restart(s)} & \makecell[l]{Time to start\\container restart(s)} \\
        \hline
        Default Probes & $8.7\pm0.5$ & $38.9\pm0.3$ \\
        Fast Probes & $0.7\pm0.5$ & $\hphantom{3}2.9\pm0.5$ \\
        \makecell[l]{SKI} & $0.1\pm0.3$ & $\hphantom{3}2.2\pm0.4$ \\
    \end{tabular}
\end{table}

\par

\begin{figure*}[!t]
\centering
\includegraphics[width=\textwidth]{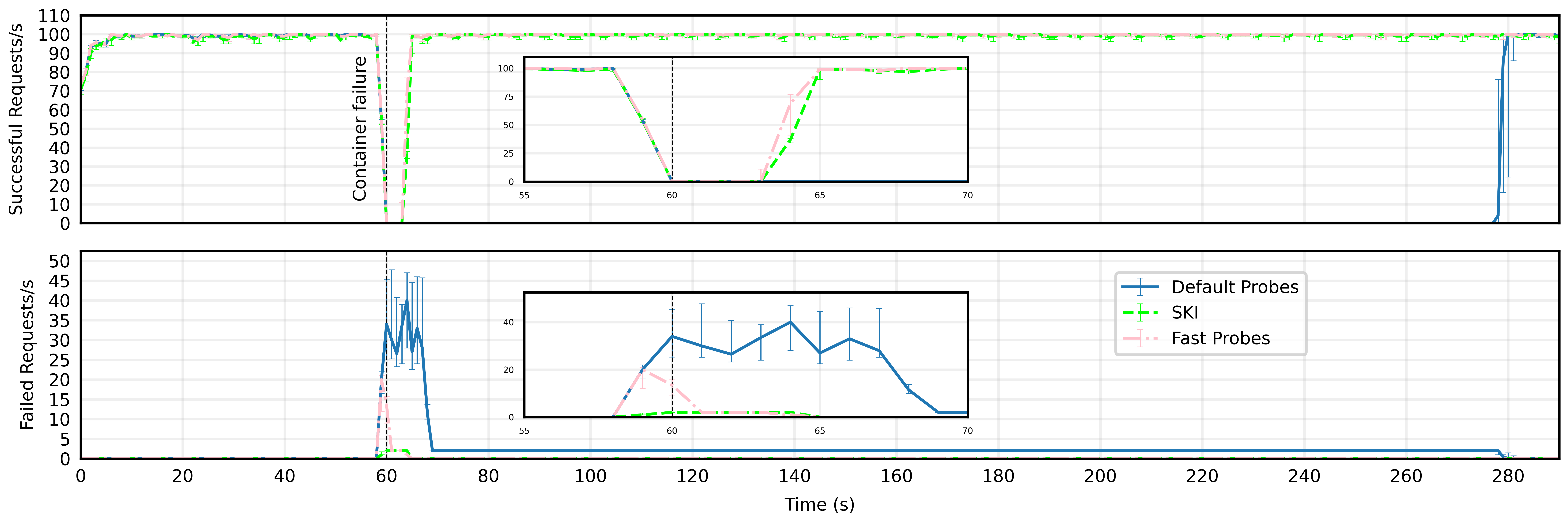}
\caption{The median successful and failed requests, with upper and lower quartiles, of the \ms{catalogue} under default probes (blue), SKI (green), and fast probes (pink) when evaluating liveness. The container recovers in 7s under SKI and fast probes.}
\label{fig:liveness_throughput}
\end{figure*}

\par

Analogous to \cref{fig:readiness_throughput}, \cref{fig:liveness_throughput} shows the median throughput and failed requests of \ms{the catalogue} during liveness evaluations.
With \ts{DP} \ms{the catalogue} recovers after $\sim$220s because of an 180s readiness probe delay, plus the 38.9s to restart the container.
The inset graph shows a peak in failed requests under \ts{DP} between $\sim$60s and $\sim$67s, when the container has failed but is still marked as \ready{} in Kubernetes.
\emph{\ts{SKI} and \ts{FP} recover the container in just 7s}, when throughput returns to $\sim$100 requests/s.
The rate of failed requests peaks at 20 requests/s under \ts{FP}, 18 (900\%) higher than under \ts{SKI}, which we attribute to the longer failure detection time.

\par

As a measure of the impact of failure detection time, \cref{tab:liveness_requests} shows the total requests sent, the total failed, and those that failed with an HTTP 500 Internal Server Error, the remaining failed requests time out. 
The large standard deviation of \ts{FP} and an inflated mean are attributed to additional container restarts (\cref{sub:additional_restarts}).
The prolonged peak of failed requests under \ts{DP} in \cref{fig:liveness_throughput} suggests HTTP 500 responses occur when the container has failed, but the container is still marked as \cmmnt{\healthy{} and} \ready{} in Kubernetes.
During that time, requests will be sent to a container that cannot process them, so a portion of these responses are still expected if SockShop is deployed with multiple Pods per service.
Considering container restarts while requests were generated, there are 30$\pm$25 HTTP 500 responses per container restart under \ts{FP}.
\emph{There are 1.2 HTTP 500 responses per container restart under \ts{SKI}, 29 (96\%) fewer than \ts{FP}.
There are 11.4 total failed requests under \ts{SKI}, 8.6 (43\%) fewer than under readiness \ts{FP} (\cref{tab:readiness_requests}), when CRI-O detects the failure.}

\par

\begin{table}[!t]
\centering
    \caption{Total requests to \ms{the catalogue}, total failed requests, and failed requests that returned an HTTP 500 response.}
    \label{tab:liveness_requests}
    \begin{tabular}{@{}llll@{}} 
        \makecell[l]{Monitoring\\Configuration} & \makecell[l]{Total \\Requests} & \makecell[l]{Total Failed\\Requests} & \makecell[l]{HTTP 500\\Responses} \\
        \hline
        Default Probes & $7600\hphantom{0}\pm100$ & $710\hphantom{.4}\pm40$ & $280\hphantom{.2}\pm50$ \\
        Fast Probes & $27000\pm2000$ & $100\hphantom{.4}\pm70$ & $\hphantom{2}60\hphantom{.2}\pm40$ \\
        \makecell[l]{SKI} & $28000\pm1000$ & $\hphantom{2}11.4\pm0.5$ & $\hphantom{28}1.2\pm0.4$ \\
    \end{tabular}
\end{table}

\par

\paragraph*{Summary}
Throughput recovers in $\sim$7s under \ts{SKI} and \ts{FP} compared to $\sim$220s under \ts{DP}.
\ts{SKI} detects failure in 0.1s, 86\% faster than \ts{FP}.
Under \ts{SKI} there are 96\% fewer HTTP 500 responses per container restart than \ts{FP}.
Some HTTP 500 responses are still expected with more than 1 Pod per service.
Under \ts{SKI}, there are 11.4 total failed requests per restart, 43\% fewer than when CRI-O detects a failure.

\subsection{Additional Container Restarts}
\label{sub:additional_restarts}

Due to the \ms{catalogue} bug identified in \cref{sect:other_container_orchestrators}, one container restart is expected when using \ts{DP} and \ts{SKI} and two when using \ts{FP}.
\emph{\con{the catalogue} was restarted between 2 and 6 times with a mean of $3\pm1$ restarts under \ts{FP}.
The kubelet logs show that the additional restarts are due to failing liveness probes.}

\par

\begin{figure*}[!t]
\centering
\includegraphics[width=\textwidth]{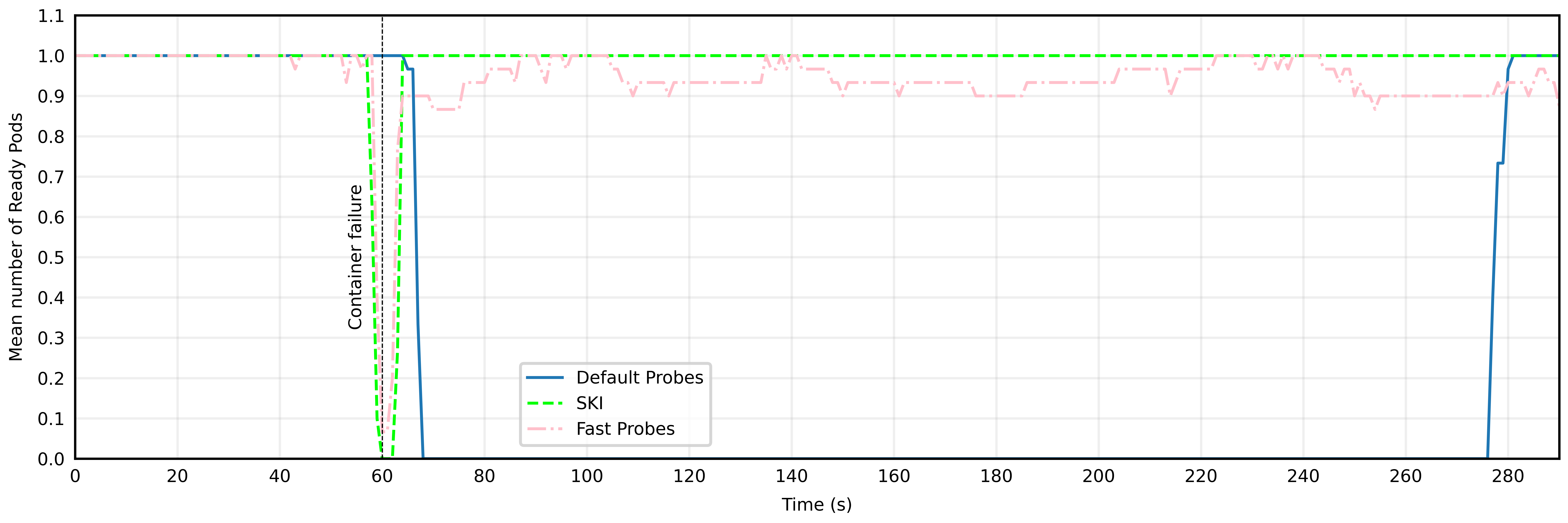}
\caption{The mean number of \ready{} \pd{catalogue}s under default probes (blue), SKI (green), and fast probes (pink).}
\label{fig:catalogue_readiness}
\end{figure*}

\cref{fig:catalogue_readiness} plots the mean number of \ready{} \pd{catalogue}s over 290s with a container failure at 60s.
Because SockShop runs with one Pod per service, a mean \cmmnt{number of \ready{} Pods} below 1 indicates when \ms{the catalogue} in a portion of tests was unavailable.
It remains unclear why most restarts under \ts{FP} occur after the failure at 60s.
To quantify availability, \cref{tab:availability} presents the cumulative mean of \ready{} Pods over time.
Availability under \ts{FP} and \ts{SKI} is 3.4 and 3.6 times greater than under \ts{DP}.
\emph{Availability under \ts{FP} is 274 Pods, 12 Pods (4\%) lower than \ts{SKI}, which we attribute to the additional container restarts.}

\par

\begin{table}[!t]
\centering
    \caption{Cumulative count of \ready{} catalogue Pods in \cref{fig:catalogue_readiness} over 290s. Under 'No Failure', one Pod is always \ready{}.}
    \label{tab:availability}
    \begin{tabular}{@{}lll@{}} 
        \makecell[l]{Monitoring\\Configuration} & \makecell[l]{Availability\\(Pods)} & \makecell[l]{Decrease Relative\\to No Failure} \\
        \hline
        No Failure & $291$ & $\hphantom{21}0$ ($\hphantom{7}0\%$) \\
        Default Probes & $\hphantom{2}80$ & $211$ ($73\%$) \\
        Fast Probes & $274$ & $\hphantom{2}17$ ($\hphantom{7}6\%$) \\
        \makecell[l]{SKI} & $286$ & $\hphantom{21}5$ ($\hphantom{7}2\%$) \\
    \end{tabular}
\end{table}

\par

Availability under \ts{FP} should be even lower due to CrashLoopBackOff delaying container restarts, but \emph{delays were inconsistently applied.}
CrashLoopBackOff delays start at 10s for the third container restart and double each subsequent restart up to a maximum of 5 minutes.
We hypothesise there is a race condition where the kubelet does not delay a restart for short termination grace periods.

\par

\paragraph*{Summary}
Under \ts{FP}, \con{the catalogue} restarts on average $3\pm1$ times, \cmmnt{above the expected 2,} leading to availability of 274 Pods, 12 Pods (4\%) lower than \ts{SKI}.
This result shows why PCM must be tuned carefully, as availability will be higher under probes that are slower to recover a failed container but do not erroneously detect failures.
However, there will be more failed requests per restart under such probes (\cref{sub:faster_liveness}).

\subsection{Discussion}
\label{sub:eval_critique}

A 1-node cluster with 1 Pod per service is used because \ts{SKI} containers only monitor 1 Pod each and do not coordinate.
When the number of Pods increases, the latency of fetching container logs via the Kubernetes API may increase.
This could be mitigated by reading logs through a shared volume (\cref{sub:scmk_alternatives}).

\par

Our evaluation uses a simple failure model:
The \prog{catalogue} is terminated once at a consistent time, not in bursts or randomly according to a probability distribution.
SockShop has few services compared to industry applications, but we believe it is adequate for tests using a simple failure model.
However, \emph{the evaluation is sufficient to show that SKI detects failure faster than probes (\cref{sub:faster_liveness}), demonstrate the need to carefully tune probes (\cref{sub:additional_restarts}),  and validate models for liveness and readiness detection times (\cref{sect:scm}).}

\section{Related Work}
\label{sect:related_work}

Silva et al. present a taxonomy of 117 microservice faults categorised by non-functional requirement \cite{silva_microservice_fault_taxonomy}.
Zhou et al. report and categorise 22 microservice faults from an industry survey \cite{zhou_fault_survey}.
The authors replicate all 22 faults using the TrainTicket microservice benchmark \cite{trainticket}.

\par

Flora et al. found Kubernetes probes failed to detect a subset of faults from \cite{zhou_fault_survey}, including errors in business logic and long response times \cite{flora_kubernetes_fault_tolerance}.
The authors concluded that probes detect faults that lead to the container failing to respond to requests but are insufficient when detecting other faults.

\par

SKI and Kubernetes probes detect container failure and then take remedial action.
Another approach to microservice fault tolerance is to predict and mitigate faults before they occur.
For example, \cite{zhou_error_prediction} uses a machine learning model trained on system logs to predict faults.

\par

We study signal-based monitoring in microservices, but the concept is present in other distributed systems.
Actor languages and frameworks like Erlang~\cite{armstrong_thesis}, Elixir~\cite{elixir}, and the Akka framework~\cite{akka}, detect failure using signals.
When an actor fails, the runtime notifies a supervisor that restarts it.

\par

The systemd \cite{systemd} service manager supports signal-based monitoring through the sd\_notify library.
Managed services may signal their state, e.g. if they are reloading configuration.
Systemd offers a watchdog feature that marks a service failed if the time between two keep-alive pings is too long.
Similar behaviour may be beneficial in detecting deadlocked container applications with SCM as they would be unable to emit signals.
\ptcom{Worth remarking that these are heartbeats, also present in Actor VMs?}
\jrcom{Maybe worth a sentence if there's enough space saved}

\section{Conclusion}
\label{sect:conclusions}

Availability is critical for many microservice applications, and container orchestrators must promptly detect failure and readiness to maximise it.
Under Poll-based Container Monitoring (PCM), an orchestrator periodically polls the state of containers.
PCM requires careful hardware and application-specific tuning to minimise the time to detect state changes, reduce resource usage, and maintain availability.
Under Signal-based Container Monitoring (SCM), containers signal the orchestrator when they change state.
SCM does not require careful tuning, but containers must emit appropriate signals.
SCM detects failure and readiness faster than PCM, as signal latency is generally far less than polling intervals.
A survey of six popular orchestrators shows that only Azure Service Fabric supports a form of SCM, which requires container modifications (\cref{sect:other_container_orchestrators}).
We offer an example where the PCM used by three orchestrators leads to container failure.

\par

We develop and empirically validate mathematical models for SCM and PCM liveness and readiness detection times (\cref{sect:scm}).
We outline the design and implementation of an implementation of SCM for the Kubernetes orchestrator, \ts{SKI} (\cref{sect:scmk}).
We compare the service availability of the SockShop benchmark under \ts{SKI} and Kubernetes probes.
To our knowledge, this is the first comparison of poll-based and signal-based monitoring for microservices.
\ts{SKI} increases Pod CPU use by 0.04 vCPU (36\%) and working set size by 9MiB (129\%) over the \ts{DP} and \ts{FP} probe configurations (\cref{sub:overheads}).
We discuss how a more sophisticated SCMK implementation could reduce these overheads.
\ts{SKI} monitoring did not decrease the throughput of the monitored service (\cref{fig:readiness_throughput,fig:liveness_throughput}).
\ts{SKI} detects \cmmnt{container} failure in 0.1s, 99\% faster than Sockshop default probes (DP) and 86\% faster than tuned fast probes (FP) (\cref{tab:failure_detection_times}).
\ts{SKI} and FP detect readiness after a failure in 2.9s and 2.2s, significantly faster than the $\sim$180s under DP (\cref{sub:slower_readiness}).
Poorly tuned probes may falsely detect \cmmnt{container} failure, leading to 4\% lower service availability compared to \ts{SKI} (\cref{fig:catalogue_readiness}).
We propose that orchestrators offer SCM features to detect failures faster and increase service availability without the drawbacks of PCM.

\paragraph*{Future Work}
\label{sub:future}

We plan to evaluate \ts{SKI} against more sophisticated fault models, and using additional benchmarks, e.g. TrainTicket~\cite{trainticket} and TeaStore~\cite{teastore}.
We intend to use Chaos Mesh \cite{chaos_mesh} to inject network latency into communications to emulate slow HTTP responses.
We expect PCM to erroneously detect high network latency as container failure, unlike \ts{SKI}.

\par

We are investigating improvements to the architecture of \ts{SKI}, so Kubernetes is not used to fetch container logs (\cref{sub:scmk_alternatives}).
Updating the kubelet to notify \ts{SKI} when a container restarts may be viable instead of polling Kubernetes.
We anticipate that both will reduce the time to detect \cmmnt{container} readiness by reducing communication. 

\noindent
\textbf{Acknowledgments}
This work was funded in part by the UK EPSRC grants STARDUST (EP/T014628) and an Amazon Research Award on Automated Reasoning.

\bibliography{IEEEabrv,bib.bib}

\begin{thebibliography}{10}
\providecommand{\url}[1]{#1}
\csname url@samestyle\endcsname
\providecommand{\newblock}{\relax}
\providecommand{\bibinfo}[2]{#2}
\providecommand{\BIBentrySTDinterwordspacing}{\spaceskip=0pt\relax}
\providecommand{\BIBentryALTinterwordstretchfactor}{4}
\providecommand{\BIBentryALTinterwordspacing}{\spaceskip=\fontdimen2\font plus
\BIBentryALTinterwordstretchfactor\fontdimen3\font minus \fontdimen4\font\relax}
\providecommand{\BIBforeignlanguage}[2]{{%
\expandafter\ifx\csname l@#1\endcsname\relax
\typeout{** WARNING: IEEEtran.bst: No hyphenation pattern has been}%
\typeout{** loaded for the language `#1'. Using the pattern for}%
\typeout{** the default language instead.}%
\else
\language=\csname l@#1\endcsname
\fi
#2}}
\providecommand{\BIBdecl}{\relax}
\BIBdecl

\bibitem{microservices}
J.~Lewis and M.~Fowler, ``Microservices,'' \url{https://martinfowler.com/articles/microservices.html}.

\bibitem{silva_microservice_fault_taxonomy}
F.~Silva, V.~Lelli, I.~Santos, and R.~Andrade, ``Towards a fault taxonomy for microservices-based applications,'' in \emph{Brazilian Symp. on Software Engineering (SBES))}, Virtual Event, Brazil, 2022.

\bibitem{zhou_fault_survey}
X.~Zhou, X.~Peng, T.~Xie, J.~Sun, C.~Ji, W.~Li, and D.~Ding, ``Fault analysis and debugging of microservice systems: Industrial survey, benchmark system, and empirical study,'' \emph{IEEE Transactions on Software Engineering}, vol.~47, no.~2, 2021.

\bibitem{zhou_error_prediction}
X.~Zhou, X.~Peng, T.~Xie, J.~Sun, C.~Ji, D.~Liu, Q.~Xiang, and C.~He, ``Latent error prediction and fault localization for microservice applications by learning from system trace logs,'' in \emph{European Software Engineering Conf. (ESEC/FSE)}, Tallinn, Estonia, 2019.

\bibitem{kubernetes}
``Kubernetes,'' https://kubernetes.io/.

\bibitem{marathon}
``Marathon,'' \url{https://github.com/d2iq-archive/marathon/tree/master}.

\bibitem{azure_sf}
``Azure service fabric,'' \url{https://azure.microsoft.com/en-gb/products/service-fabric/}.

\bibitem{casalicchio_container_orchestration}
E.~Casalicchio, \emph{Container Orchestration: A Survey}.\hskip 1em plus 0.5em minus 0.4em\relax Springer International Publishing, 2019, pp. 221--235.

\bibitem{sockshop}
Weaveworks, ``Sockshop,'' \url{https://github.com/microservices-demo/microservices-demo/tree/master}.

\bibitem{monzo_deployment_rate}
\BIBentryALTinterwordspacing
How we deploy to production over 100 times a day. [Online]. Available: \url{https://monzo.com/blog/2022/05/16/how-we-deploy-to-production-over-100-times-a-day}
\BIBentrySTDinterwordspacing

\bibitem{datadog_container_report}
\BIBentryALTinterwordspacing
10 insights on real-world container use. [Online]. Available: \url{https://www.datadoghq.com/container-report/}
\BIBentrySTDinterwordspacing

\bibitem{heorhiadi_gremlin}
V.~Heorhiadi, S.~Rajagopalan, H.~Jamjoom, M.~K. Reiter, and V.~Sekar, ``Gremlin: Systematic resilience testing of microservices,'' in \emph{Int. Conf. on Distributed Computing Systems (ICDCS)}, Nara, Japan, 2016.

\bibitem{sahoo_failure_analysis}
R.~Sahoo, M.~Squillante, A.~Sivasubramaniam, and Y.~Zhang, ``Failure data analysis of a large-scale heterogeneous server environment,'' in \emph{Int. Conf. on Dependable Systems and Networks (DSN)}, Florence, Italy, 2004.

\bibitem{garraghan_cloud_failure_analysis}
P.~Garraghan, P.~Townend, and J.~Xu, ``An empirical failure-analysis of a large-scale cloud computing environment,'' in \emph{Int. Symp. on High Assurance Systems Engineering (HASE)}, Miami Beach, FL, USA, 2014.

\bibitem{crio}
``Cri-o,'' https://cri-o.io/.

\bibitem{straesser_container_start_times}
M.~Straesser, A.~Bauer, R.~Leppich, N.~Herbst, K.~Chard, I.~Foster, and S.~Kounev, ``An empirical study of container image configurations and their impact on start times,'' in \emph{Int. Symp. on Cluster, Cloud and Internet Computing (CCGrid)}, Bangalore, India, 2023.

\bibitem{gitlab_kubernetes_fault}
S.~Azzopardi, ``How we reduced 502 errors by caring about pid 1 in kubernetes,'' https://about.gitlab.com/blog/2022/05/17/how-we-removed-all-502-errors-by-caring-about-pid-1-in-kubernetes/.

\bibitem{doordash_kubernetes_fault}
A.~Ivanov, ``How to handle kubernetes health checks,'' \url{https://doordash.engineering/2022/08/09/how-to-handle-kubernetes-health-checks/}.

\bibitem{aws_eks}
``Amazon elastic kubernetes service,'' \url{https://aws.amazon.com/eks/}.

\bibitem{azure_aks}
``Azure kubernetes service,'' \url{https://azure.microsoft.com/en-gb/products/kubernetes-service}.

\bibitem{gcp_gke}
``Google kubernetes engine,'' \url{https://cloud.google.com/kubernetes-engine}.

\bibitem{fabricobserver}
``Fabricobserver,'' \url{https://github.com/microsoft/service-fabric-observer}.

\bibitem{nomad}
``Nomad,'' https://www.nomadproject.io/.

\bibitem{breck_probes_presentation}
``Kubernetes probes: How to avoid shooting yourself in the foot by colin breck,'' \url{https://www.youtube.com/watch?v=UIHfzUvMAbM}.

\bibitem{docker}
``Docker,'' \url{https://www.docker.com/}.

\bibitem{aws_ecs}
``Amazon elastic container service,'' https://aws.amazon.com/ecs/.

\bibitem{kubernetes_toolkit_patterns}
``The distributed system toolkit: Patterns for composite containers,'' \url{https://kubernetes.io/blog/2015/06/the-distributed-system-toolkit-patterns/}.

\bibitem{prometheus}
``Prometheus,'' {https://prometheus.io/}.

\bibitem{locust}
``Locust,'' https://locust.io/.

\bibitem{aws_ms_ec2_recommendation}
``New seventh-generation general purpose amazon ec2 instances (m7i-flex and m7i),'' \url{https://aws.amazon.com/blogs/aws/new-seventh-generation-general-purpose-amazon-ec2-instances-m7i-flex-and-m7i/}.

\bibitem{gitlab_kubernetes_merge}
R.~Coutable, ``Set a sane default of 2 seconds for workhorse 'readinessprobe.timeoutseconds','' \url{https://gitlab.com/gitlab-org/charts/gitlab/-/merge_requests/964}.

\bibitem{trainticket}
``Trainticket,'' \url{https://github.com/FudanSELab/train-ticket/}.

\bibitem{flora_kubernetes_fault_tolerance}
J.~Flora, P.~Gonçalves, M.~Teixeira, and N.~Antunes, ``A study on the aging and fault tolerance of microservices in kubernetes,'' \emph{IEEE Access}, vol.~10, pp. 132\,786--132\,799, 2022.

\bibitem{armstrong_thesis}
J.~Armstrong, ``Making reliable distributed systems in the presence of software errors,'' Dec 2003.

\bibitem{elixir}
``Elixir,'' https://elixir-lang.org/.

\bibitem{akka}
``Akka,'' https://akka.io/.

\bibitem{systemd}
``Systemd,'' \url{https://github.com/systemd/systemd}.

\bibitem{teastore}
J.~von Kistowski, S.~Eismann, N.~Schmitt, A.~Bauer, J.~Grohmann, and S.~Kounev, ``{TeaStore: A Micro-Service Reference Application for Benchmarking, Modeling and Resource Management Research},'' in \emph{Int. Symp. on the Modelling, Analysis, and Simulation of Computer and Telecommunication Systems (MASCOTS)}, Milwaukee, WI, USA, 2018.

\bibitem{chaos_mesh}
``Choas mesh,'' https://chaos-mesh.org/.

\end{thebibliography}

\end{document}